\documentclass[twoside]{article}
\usepackage{qic,epsfig}
\usepackage{amsmath}
\usepackage{amsfonts}
\usepackage{amssymb}
\usepackage{graphicx}
\usepackage{cite}
\usepackage{enumerate}

\usepackage{color}

	\newcommand{\ket}[1]{\left| #1 \right\rangle}
	\newcommand{\bra}[1]{\left\langle #1 \right|}

\begin{document}
\setlength{\textheight}{8.0truein}  

\runninghead{The entangling power of a ``glocal" dissipative map}
            {A. Nourmandipour, M.K. Tavassoly, S. Mancini}

\normalsize\textlineskip
\thispagestyle{empty}
\setcounter{page}{1}

\copyrightheading{0}{0}{2003}{000--000}

\vspace*{0.88truein}

\alphfootnote

\fpage{1}

\centerline{\bf 
The entangling power of a ``glocal" dissipative map}
\vspace*{0.37truein}
\centerline{\footnotesize
ALIREZA NOURMANDIPOUR}
\vspace*{0.015truein}
\centerline{\footnotesize\it Atomic and Molecular Group, Faculty of Physics,
}
\baselineskip=10pt
\centerline{\footnotesize\it Yazd University, Yazd  89195-741, Iran}
\baselineskip=10pt
\centerline{\footnotesize\it anoormandip@stu.yazd.ac.ir}

\vspace*{10pt}
\centerline{\footnotesize 
M. K. TAVASSOLY}
\vspace*{0.015truein}
\centerline{\footnotesize\it Atomic and Molecular Group, Faculty of Physics,
}
\baselineskip=10pt
\centerline{\footnotesize\it Yazd University, Yazd  89195-741, Iran}

\vspace*{0.015truein}
\centerline{\footnotesize\it The Laboratory of Quantum Information Processing,
}
\baselineskip=10pt
\centerline{\footnotesize\it Yazd University, Yazd  89195-741, Iran}
\baselineskip=10pt
\centerline{\footnotesize\it mktavassoly@yazd.ac.ir}
\vspace*{10pt}
\centerline{\footnotesize 
STEFANO MANCINI}
\vspace*{0.015truein}
\centerline{\footnotesize\it School of Science \& Technology,
}
\baselineskip=10pt
\centerline{\footnotesize\it University of Camerino, I-62032 Camerino, Italy}

\vspace*{0.015truein}
\centerline{\footnotesize\it INFN-Sezione di Perugia, Via A. Pascoli,
}
\baselineskip=10pt
\centerline{\footnotesize\it I-06123 Perugia, Italy}
\baselineskip=10pt
\centerline{\footnotesize\it stefano.mancini@unicam.it}
\vspace*{0.225truein}
\publisher{(received date)}{(revised date)}

\vspace*{0.21truein}

\abstracts{
We consider a model of two qubits dissipating into both local and global environments (generally at non-zero temperatures), with the possibility of interpolating between purely local dissipation and purely global one.  
The corresponding dissipative dynamical map is characterized in terms of its Kraus operators focusing on the stationary regime. 
We then determine conditions under which entanglement can be induced by the action of such a map. It results (rather counterintuitively) that in order to have entanglement in the presence of local environment, this latter must be at nonzero temperature.
}{}{}

\vspace*{10pt}

\keywords{Open quantum systems, Entanglement measures}
\vspace*{3pt}
\communicate{to be filled by the Editorial}

\vspace*{1pt}\textlineskip   

\section{Introduction}

Entanglement is one of the most interesting features of quantum mechanics that cannot be explained by any local classical theory. This notion plays a key role in quantum information and quantum computation sciences (see e.g. \cite{Nielsen2010}). As soon as entanglement was recognized as a resource for quantum information processing it was considered very fragile and easily degradable by environmental noise. So the idea to avoid as much as possible interactions with environment was dominant. However, recently it has been put forward the idea that environment can after all have a positive role \cite{VWC09}. For instance, when environment acts globally on a composite system it can supply a kind of interaction that helps in establishing entanglement.
Among environmental effects, dissipation plays a prominent role because it allows for the stabilization of targeted resources, like entanglement, a fact that may result as a key advantage over unitary (noiseless) manipulation \cite{Kraus2008}.
Then, it has been shown, both theoretically \cite{Plenio2002} and experimentally \cite{Krauter2011}, that a global dissipative environment can establish stationary entanglement.
Surprisingly, this happens even without any direct interaction among subsystems \cite{Ghosh2006}. The simplest model where such an effect occurs is that of two qubits dissipating into a common environment. A possibility that has been proved true for systems composed by more than two subsystems \cite{Memarzadeh2013}.

After having ascertained the benefits of global environment's action on entanglement, 
one is naturally led to ask what would happen if beside there are also local environments actions.
Would entanglement  be generated as well and persist indefinitely? 
If yes, to what amount and for which initial states?
Here, we shall address these issues by considering
a model of two qubits dissipating into a ``glocal" environment (at non-zero temperatures). 
By ``glocal" we mean a mixture of global environment (with which the two qubits jointly interact) and local environments (with which each qubit separately interacts).

We shall then determine conditions under which stationary entanglement can be induced. It results on the one hand (and rather counterintuitively) that entanglement in presence of local environments is achievable when these are at nonzero temperature.
On the other hand, while global environment is vital for indirect qubits interaction, it should be ideally at zero temperature.

The results are obtained by first studying the dynamical map (focusing on the stationary regime) in terms of its Kraus operators \cite{Kraus83} and then by characterizing it through an entangling power measure. This latter relies on the statistical average over the initial states establishing an input-independent dynamics of entanglement \cite{Zanardi2000} (a concept that has already been applied in many quantum systems \cite{Lakshminarayan2001}).

The paper is organized as follow. In Sec. \ref{sec:model} we introduce our model with two qubits and thermal environments. In Sec. \ref{sec:Kraus} we find the Kraus operators corresponding to the dynamical map focusing on the stationary regime. Sec. \ref{sec:entpower} deals with the entangling power of the map. Finally, we draw our conclusions in Sec. \ref{sec:conclusion}.


\section{The Model}\label{sec:model}

\begin{figure}[h]
\centering
\includegraphics[width=0.45\textwidth]{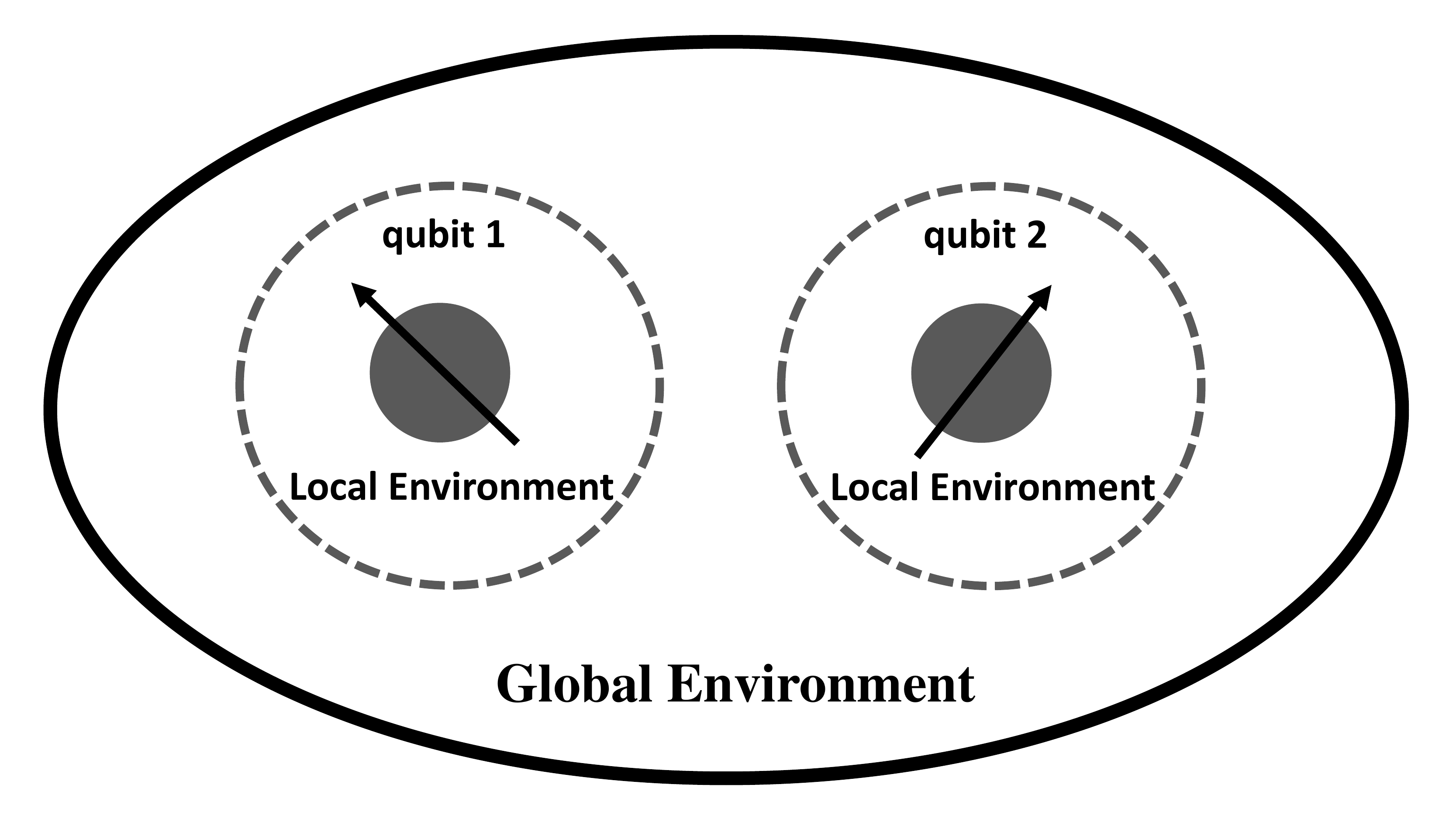}
\fcaption{\label{fig:sys} Pictorial representation of the system under study.}
\end{figure}

Dissipation of energy into environment is an important phenomenon in a variety of open quantum systems. Quite generally, the environment should be treated as a distribution of the uncorrelated thermal equilibrium mixture of states. For two qubits dissipating into their own thermal environments, the description of the dynamics stems on a master equation of Lindblad form  \cite{Breuer2002}, with Lindblad operators proportional to $\sigma_1$, $\sigma_2$, $\sigma_1^{\dagger}$ and $\sigma_2^{\dagger}$ respectively, where $\sigma_i:=| g_i\rangle\langle e_i|$, being $|g_i\rangle$, $|e_i\rangle$ the ground and the excited state respectively of the $i$th qubit $(i=1,2)$.
This dynamics constitutes the local dissipation.
Here, driven by the fact that the continuous miniaturization of physical devices makes qubits closely spaced, we are going to also consider global dissipation, namely the two qubits dissipating into a  common thermal environment (see Fig. \ref{fig:sys}). 
This amounts to consider additional Lindblad operators proportional 
to $\sigma_1 + \sigma_2$ and $\sigma_1^{\dagger}+\sigma_2^{\dagger}$. 
Thus the dynamics of the density operator $\rho$ of the system under study will be governed by the following master equation
\begin{equation}
\dot{\rho}(t)=\gamma \sum_{k=1}^2 \left[2L_k\rho L_k^\dag
-L_k^\dag L_k \rho -\rho L_k^\dag L_k\right]+(1-\gamma) \sum_{k=3}^6 \left[2L_k\rho L_k^\dag
-L_k^\dag L_k \rho -\rho L_k^\dag L_k\right],
\label{eq:newsys}
\end{equation}
where
\begin{equation}
\begin{aligned}
L_1&=\sqrt{\bar{n}_g+1} \;(\sigma_1+\sigma_2),\\
L_2&=\sqrt{\bar{n}_g} \;(\sigma_1^\dag+\sigma_2^\dag),\\
L_3&=\sqrt{\bar{n}_l+1}\; \sigma_1,\\
L_4&=\sqrt{\bar{n}_l+1}\;\sigma_2,\\
L_5&=\sqrt{\bar{n}_l}\;\sigma_1^\dag,\\
L_6&=\sqrt{\bar{n}_l}\;\sigma_2^\dag,
\end{aligned}
\label{eq:Lin}
\end{equation}
with $\bar{n}_g$ and $\bar{n}_l$ the number of thermal excitations in the global and local environments. 
In Eq.\eqref{eq:newsys} the parameter $\gamma\in[0,1]$ describes the interplay between purely local dissipation ($\gamma=0$) and purely global dissipation ($\gamma=1$). We have assumed a unit decay rate. 

In case $\bar{n}_l=\bar{n}_g=0$, since in
Eq.\eqref{eq:newsys} there is no Hamiltonian term and the Lindblad operators all commute, we can just model the process as a ``weakly measure and prepare" channel. The local dissipation just asks each qubit whether they are excited and give them some chance of decaying in the ground state. It can be represented by a Markov link 
$|e_k\rangle\to|g_k\rangle$ $(k=1,2)$. The global dissipation just asks the two qubits 
for the presence of one or two excitations in the symmetric subspace, which then decay with a fixed rate.
 It leaves $\frac{1}{\sqrt{2}}\left(\ket{e_1}\ket{g_2}-\ket{g_1}\ket{e_2}\right )$ fixed and it can be represented by a Markov link 
$\ket{e_1}\ket{e_2}\to\frac{1}{\sqrt{2}}\left(\ket{e_1}\ket{g_2}+\ket{g_1}\ket{e_2}\right )\to\ket{g_1}\ket{g_2}$. In case of nonzero $\bar{n}_l$ and or $\bar{n}_g$, the dynamics results much more involved. 

To study it we formally expand the density operator in the basis $\{\ket{1}:=\ket{e_1}\ket{e_2},\ket{2}:=\ket{e_1}\ket{g_2},\ket{3}:=\ket{g_1}\ket{e_2},\ket{4}:=\ket{g_1}\ket{g_2}\}$ so to have
\begin{equation}
\rho(t)= \sum_{j,k=1}^{4}\rho_{jk}(t)\ket{j}\bra{k},
\label{eq:timeden}
\end{equation}
where $\rho_{jk}(t)$ are unknown time-dependent coefficients.
For the sake of simplicity we will define
$\rho_{jk}\equiv \rho_{jk}(0)$.

Upon insertion of \eqref{eq:timeden} into \eqref{eq:newsys} 
the dynamics will be described by a set of linear differential equations for the unknown coefficients
$\rho_{jk}(t)$ that can be compactly expressed as
\begin{equation}
\dot{\text{\bf{v}}}(t)=M\text{\bf{v}}(t),
\label{eq:differ}
\end{equation}
where $\text{\bf{v}}(t)=\left( \rho_{11}(t),\rho_{12}(t),\cdots,\rho_{43}(t),\rho_{44}(t)\right) ^{\text{T}}$ and $M$ is a $16\times 16$ matrix of constant coefficients given by
\begin{equation}
M= \begin{pmatrix}
  M_{11} & M_{12} \\
  M_{21} & M_{22}
  \end{pmatrix},
\end{equation}
with
\begin{equation}
M_{11}:=  \left(
  \begin{matrix}
    -4 & 0 & 0 & 0 & 0 & 2\xi & \eta & 0\\ 
    0 & -3 & \zeta & 0 & 0 & 0 & 0 & \eta \\ 
    0 & \zeta & -3 & 0 & 0 & 0 & 0 & 2\xi \\ 
    0 & 0 & 0 & -2 & 0 & 0 & 0 & 0 \\
    0 & 0 & 0 & 0 & -3 & 0 & 0 & 0 \\
    2(1+\xi) & 0 & 0 & 0 & 0 & -2 & \zeta & 0 \\
    \chi& 0 & 0 & 0 & 0 & \zeta & -2 & 0  \\
     0 & \chi& 2(1+\xi) & 0 & 0 & 0 & 0 & -1 
  \end{matrix}\right)-4\xi\, I_{8\times 8},
\end{equation}

\begin{equation}
M_{12}:=  \left(
  \begin{matrix}
 0 & \eta & 2\xi & 0 & 0 & 0 & 0 & 0 \\
    0 & 0 & 0 & 2\xi & 0 & 0 & 0 & 0 \\
    0 & 0 & 0 & \eta & 0 & 0 & 0 & 0 \\
    0 & 0 & 0 & 0 & 0 & 0 & 0 & 0 \\
    \zeta & 0 & 0 & 0 & 0 & \eta & 2\xi & 0 \\
    0 & \zeta & 0 & 0 & 0 & 0 & 0 & 2\xi \\
    0 & 0 & \zeta & 0 & 0 & 0 & 0 & \eta \\
    0 & 0 & 0 & \zeta & 0 & 0 & 0 & 0 
  \end{matrix}\right),
\end{equation}

\begin{equation}
M_{21}:=  \left(
  \begin{matrix}
     0 & 0 & 0 & 0 &  \zeta & 0 & 0 & 0  \\
     \chi& 0 & 0 & 0 & 0 & \zeta & 0 & 0 \\
     2(1+\xi) & 0 & 0 & 0 & 0 & 0 & \zeta & 0 \\
     0 & 2(1+\xi) & \chi& 0 & 0 & 0 & 0 & \zeta \\
     0 & 0 & 0 & 0 & 0 & 0 & 0 & 0  \\
     0 & 0 & 0 & 0 & \chi& 0 & 0 & 0 \\
     0 & 0 & 0 & 0 & 2(1+\xi) & 0 & 0 & 0  \\
     0 & 0 & 0 & 0 & 0 & 2(1+\xi) & \chi& 0 
  \end{matrix}\right),
\end{equation}

\begin{equation}
M_{22}:=  \left(
  \begin{matrix}
   -3 & 0 & 0 & 0 & 0 & 2\xi &  \eta & 0 \\
    0 & -2 & \zeta & 0 & 0 & 0 & 0 & \eta \\
    0 & \zeta & -2 & 0 & 0 & 0 & 0 & 2\xi \\
    0 & 0 & 0 & -1 & 0 & 0 & 0 & 0 \\
    0 & 0 & 0 & 0 & -2 & 0 & 0 & 0 \\
    2(1+\xi) & 0 & 0 & 0 & 0 & -1 & \zeta & 0 \\
    \chi& 0 & 0 & 0 & 0 & \zeta & -1 & 0 \\
    0 & \chi& 2(1+\xi) & 0 & 0 & 0 & 0 & 0
  \end{matrix}\right)-4\xi\, I_{8\times 8},
\end{equation}
and 
\begin{equation}
\begin{aligned}
\xi&:=\gamma\bar{n}_g+(1-\gamma)\bar{n}_l,\\ 
\eta&:=2\gamma\bar{n}_g,\\
\chi&:=2\gamma(1+\bar{n}_g),\\
\zeta&:=-\gamma(1+2\bar{n}_g).
\end{aligned}
\end{equation}


\section{Steady states and dynamical map} \label{sec:Kraus}

We are interested in the stationary solutions of Eq.\eqref{eq:differ}, i.e. in $\text{\bf{v}}(t=\infty)$.
We may notice that for $\gamma<1$ or $\bar{n}_g>0$ the steady state can be simply found by solving
$M\text{\bf{v}}(t=\infty)=0$ as $\ker M$ results one dimensional.
In contrast for $\gamma=1$ and $\bar{n}_g=0$, $\ker M$ results of dimension greater than one, meaning that the steady state is not unique and will depend on the initial state.
Hence it must be derived by first solving  Eq.\eqref{eq:differ} and then taking $\lim_{t\to\infty}\text{\bf{v}}(t)$
(see Appendix \ref{App:differ}).
This different behavior should be ascribed to the fact that in Eq.\eqref{eq:newsys} when $\gamma=1$ and $\bar{n}_g=0$ there exist non-trivial operators (i.e. not multiple of the identity) commuting with the Lindblad operators \cite{Spohn}.

Taking into account both cases, the stationary density operator can be expressed (in the basis $\{|1\rangle,
|2\rangle, |3\rangle, |4\rangle\}$) as:
\begin{equation}
\rho(\infty)=\begin{pmatrix}
 B_1 & 0 & 0 & 0 \\
 0 & B_2+R_1 & D-R_1 & R_2 \\
 0 & D-R_1 & B_3+R_1 & -R_2 \\
 0 & R_2^{*} & -R_2^{*} & B_4+R_3
 \end{pmatrix},
 \label{eq:stationarystate}
\end{equation}
where
\begin{equation}
\begin{aligned}
B_1&:=\frac{1}{H}\left(1-\delta_{\gamma,1}\delta_{\bar{n}_g,0}\right)
\Big[ 2 \gamma ^2 \bar{n}_g^2-(\gamma -1) \bar{n}_l^2 (6 \gamma 
   \bar{n}_g+1)+2 \gamma  \bar{n}_g \bar{n}_l \Big(\gamma  (2 \bar{n}_g-1)+1\Big)+2
   (\gamma -1)^2 \bar{n}_l^3\Big], \\
B_2&:=\frac{1}{H}\left(1-\delta_{\gamma,1}\delta_{\bar{n}_g,0}\right)
\Big(2 \gamma  \bar{n}_g-2 (\gamma -1) \bar{n}_l+1\Big) \Big(\gamma 
   \bar{n}_g+\bar{n}_l (2 \gamma  \bar{n}_g-\gamma  \bar{n}_l+\bar{n}_l+1)\Big), \\  
B_3&:=B_2, \\ 
B_4&:=\frac{1}{H}\left(1-\delta_{\gamma,1}\delta_{\bar{n}_g,0}\right)
\Big[ 2 \gamma ^2 (\bar{n}_g-\bar{n}_l) \big(2 \bar{n}_g
   \bar{n}_l+\bar{n}_g-\bar{n}_l^2\big)+(2 \bar{n}_l+1)
      (\bar{n}_l+1)^2 \\
   &\hspace{1cm}+\gamma  (\bar{n}_l+1) \Big(\bar{n}_g (6
   \bar{n}_l+4)-\bar{n}_l (4 \bar{n}_l+1)+1\Big)\Big], \\   
D&:=\frac{\gamma}{H}\left(1-\delta_{\gamma,1}\delta_{\bar{n}_g,0}\right)
(\bar{n}_g-\bar{n}_l), \\
H&:=8\gamma ^2 (\bar{n}_g-\bar{n}_l) \big(2 \bar{n}_g
      \bar{n}_l+\bar{n}_g-\bar{n}_l^2\big)+\gamma  (2 \bar{n}_l+1)^2 (6
      \bar{n}_g-4 \bar{n}_l+1)+(2 \bar{n}_l+1)^3, \\
R_1&:=\dfrac{1}{4}\delta_{\gamma,1}\delta_{\bar{n}_g,0}\left(\rho_{22}-\rho_{23}-\rho_{32}+\rho_{33}\right), \\
R_2&:=\dfrac{1}{2}\delta_{\gamma,1} \delta_{\bar{n}_g,0} \left(\rho_{24}-\rho_{34}\right),\\
R_3&:=\dfrac{1}{2}\delta_{\gamma,1} \delta_{\bar{n}_g,0} \left(\rho_{11}+\rho_{44}
+\rho_{23}+\rho_{32}+1\right).
\end{aligned}
\label{eq:Coes}
\end{equation}
Here it is
\begin{equation*}
\delta_{\gamma,1}:=\left\{
\begin{array}{ccc}
0 & & 0\le\gamma<1 \\
1 & & \gamma=1
\end{array}\right., 
\end{equation*}
and 
\begin{equation*}
\delta_{\bar{n}_g,0}:=\left\{
\begin{array}{ccc}
0 & & \bar{n}_g >0 \\
1 & & \bar{n}_g=0
\end{array}\right..
\end{equation*}
Notice that the dependance from the initial state is shown by terms $R_1$, $R_2$, and $R_3$.

We can consider the evolution $\rho(0)\to \rho(\infty)$ 
as resulting from a (dissipative) map, namely
\begin{equation}
\rho(\infty)={\cal D}(\rho(0)).
\label{eq:dynmap}
\end{equation}
In order to find its Kraus decomposition \cite{Kraus83} we need to treat the case
$\gamma<1$ or $\bar{n}_g>0$ separately from $\gamma=1$ and $\bar{n}_g=0$.

In the former the map ${\cal D}$ has a fixed point
\begin{equation}
\rho_{_{fixed}}(\infty)=\begin{pmatrix}
 B_1 & 0 & 0 & 0 \\
 0 & B_2 & D & 0 \\
 0 & D & B_3 & 0 \\
 0 & 0 & 0 & B_4
 \end{pmatrix}, \label{eq:staf} 
\end{equation}
hence the corresponding Kraus operators can be constructed as 
\begin{equation}
K^\prime_{jl}=\sqrt{\upsilon_l} |\psi_j\rangle\langle l|, \qquad j,l=1 \cdots 4,
\label{eq:krausstaind}
\end{equation}
by means of the spectral decomposition $\rho_{_{fixed}}(\infty)=\sum_{j=1}^4\upsilon_j |\psi_j\rangle\langle\psi_j|$,
where
\begin{equation}
\begin{aligned}
\ket{\psi_1}&=\begin{pmatrix}
 1  \\
 0  \\
 0  \\
 0 
 \end{pmatrix}, \quad 
 \ket{\psi_2}=\begin{pmatrix}
  0  \\
  0  \\
  0  \\
  1 
  \end{pmatrix}, \quad
  \ket{\psi_3}=\begin{pmatrix}
   0  \\
   -1  \\
   1  \\
   0 
   \end{pmatrix}, \quad
   \ket{\psi_4}=\begin{pmatrix}
    0  \\
    1  \\
    1  \\
    0 
    \end{pmatrix},  
\end{aligned}
\label{eq:eigendecomp}
\end{equation}
and
\begin{equation}
\begin{aligned}
\upsilon_1&=B_1, \\  
\upsilon_2&=B_4, \\ 
\upsilon_3&=B_2-D, \\ 
\upsilon_4&=B_2+D.  
\end{aligned}
\label{eq:eigenvalues}
\end{equation}

In contrast, for the case $\gamma=1$ and $\bar{n}_g=0$, the stationary state 
\begin{equation}
\label{eq:dressed}
\rho_{_{ini}}(\infty):=\begin{pmatrix}
 0 & 0 & 0 & 0 \\
 0 & R_1 & -R_1 & R_2 \\
 0 & -R_1 & R_1 & -R_2 \\
 0 & R_2^{*} & -R_2^{*} & R_3
 \end{pmatrix}
\end{equation}
depends on the initial state, hence in order to find the Kraus operators of the corresponding map
we need to first obtain them for the map
\begin{equation}
\rho(t)={\cal D}_t(\rho(0)),
\label{eq:dynmapt}
\end{equation}
where the subscript $t$ emphasizes the parametrically dependence on time. 
Then take the limit $t\to\infty$.  Implementing this procedure in Appendix \ref{App:Appendixcoeff}, 
it results
\begin{equation}
\begin{aligned}
K_1^{\prime\prime}&=\dfrac{1}{2}\begin{pmatrix}
 0 & 0 & 0 & 0 \\
 0 & 1 & -1 & 0 \\
 0 & -1 & 1 & 0 \\
 0 & 0 & 0 & 2
 \end{pmatrix}, \ \ \
K_2^{\prime\prime}=\begin{pmatrix}
  0 & 0 & 0 & 0 \\
  0 & 0 & 0 & 0 \\
  0 & 0 & 0 & 0 \\
  1 & 0 & 0 & 0
  \end{pmatrix}, \\
K_3^{\prime\prime}&=\begin{pmatrix}
   0 & 0 & 0 & 0 \\
   0 & 0 & 0 & 0 \\
   0 & 0 & 0 & 0 \\
   0 & 0 & 0 & 0
   \end{pmatrix}, \ \ \
K_4^{\prime\prime}=\dfrac{1}{\sqrt{2}}\begin{pmatrix}
   0 & 0 & 0 & 0 \\
   0 & 0 & 0 & 0 \\
   0 & 0 & 0 & 0 \\
   0 & 1 & 1 & 0
   \end{pmatrix}.
\end{aligned}
\label{eq:krausinf}
\end{equation}

Taking into account both cases  \eqref{eq:krausstaind} and \eqref{eq:krausinf}, the stationary state \eqref{eq:stationarystate} can be written as
\begin{equation}
\rho(\infty)=\sum_{jl=1}^{4}K_{jl}\rho(0) K_{jl}^{\dagger},
\end{equation} 
where
\begin{equation}
K_{jl}=\left( 1-\delta_{\gamma,1}\delta_{\bar{n}_g,0}\right)K^\prime_{jl}
+\delta_{\gamma,1}\delta_{\bar{n}_g,0} \delta_{j,l}K^{\prime\prime}_j.
\end{equation}


\section{Entangling power}\label{sec:entpower}

In what follows, we use the concurrence \cite{Wootters1998} to quantify the amount of entanglement which is defined as 
\begin{equation}
E(\rho):=\max\left\lbrace 0, \sqrt{\ell_1}- \sqrt{\ell_2}- \sqrt{\ell_3}- \sqrt{\ell_4}\right\rbrace,
\label{eq:con}
\end{equation}
where $\ell_j$, $j=1,2,3,4$, are the eigenvalues (in decreasing order) of 
$\rho\left(\sigma_1^y\otimes\sigma_2^y\rho^{*}\sigma_1^y\otimes\sigma_2^y\right)$ with $\rho^*$ the complex conjugate of $\rho$ and $\sigma_k^y:=i(\sigma_k-\sigma_k^\dag)$. 

Assume the initial state of the system to be pure and factorable. Its general parametrization is
\begin{equation}
\begin{aligned}
|\psi(0)\rangle&=\left( \cos(\theta_1/2) \ket{e_1}+\sin(\theta_1/2)e^{i\varphi_1}\ket{g_1}\right)  \\
&\otimes\left( \cos(\theta_2/2) \ket{e_2}+\sin(\theta_2/2)e^{i\varphi_2}\ket{g_2}\right),
\end{aligned}
\label{eq:inifact}
\end{equation}
with $\theta_k\in\left[ 0,\pi\right] $ and $\varphi_k\in\left[ 0,2\pi\right] $ for $k=1,2$.
Then, the concurrence \eqref{eq:con} for the stationary state \eqref{eq:stationarystate} becomes
\begin{equation}\label{steadyconc}
\begin{aligned}
E&=2\left(\left|D\right|-\sqrt{B_1 B_4}+|R_1|\right)\\
&=2\left|D\right|-2\sqrt{B_1 B_4}
+\delta_{\gamma,1}\delta_{\bar{n}_g,0}\dfrac{1}{4}
\left| 1-\cos\theta_1\cos\theta_2-\cos(\varphi_1-\varphi_2)\sin\theta_1\sin\theta_2\right|,
\end{aligned}
\end{equation}
It is worth noticing that when $\gamma=1$ and $\bar{n}_g=0$ the third term contributes, while the second does not because $B_1=0$.

\begin{figure}[ht!]
\centering
\includegraphics[width=0.6\textwidth]{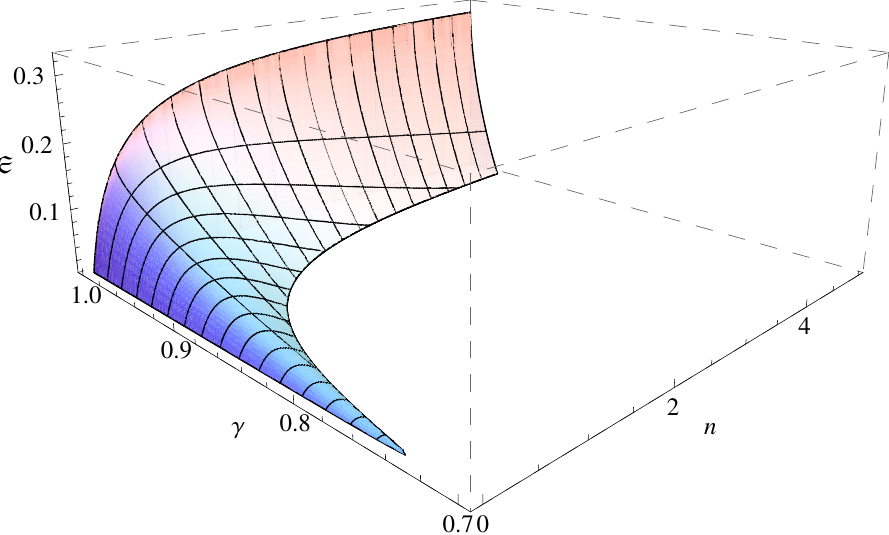}
    \fcaption{Entangling power $\mathfrak{E}$ as a function of $\gamma$ and $\bar{n}_l$ for $\bar{n}_g=0$.
    The value of $\mathfrak{E}$ for $\gamma$ exactly equal to 1 is not reported here.}
    \label{fig:entpow3D0}
\end{figure}

\begin{figure}[ht!]
\centering
\includegraphics[width=0.5\textwidth]{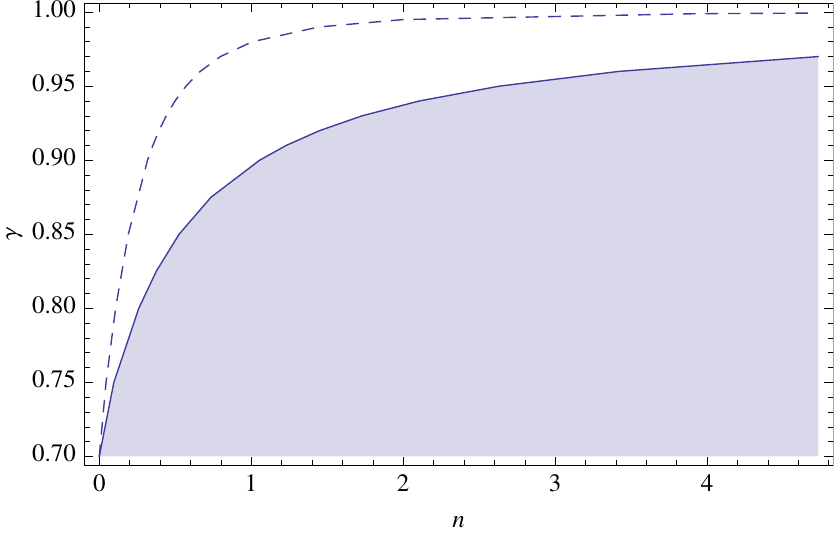}
    \fcaption{Regions of the parameter space $\{\gamma,\bar{n}_l\}$ where the entangling power $\mathfrak{E}$ is greater than zero (white) and zero (grey) for $\bar{n}_g=0$. Along the dashed line $\mathfrak E$ takes its maximum  value. }
    \label{fig:entpowregions}
\end{figure}

Quite generally the stationary entanglement \eqref{steadyconc} depends on the initial state. However, we can say that a map is a good entangler when the average of the final entanglement over all possible initial states is positive. Moving on from \cite{Zanardi2000} we define the entangling power of ${\cal D}$ as
\begin{equation}
{\mathfrak E}({\cal D}):=\int E\left( {\cal D} ( |\psi(0)\rangle\langle \psi(0)|)\right) \, d\mu( |\psi(0)\rangle),
\label{eq:enpower}
\end{equation}
where $ d\mu( |\psi(0)\rangle)$ is the probability measure over the submanifold of factorable states in $\mathbb{C}^2\otimes \mathbb{C}^2$. The latter is induced by the Haar measure of ${\rm SU}(2) \otimes {\rm SU}(2)$. Specifically, referring to the parametrization of \eqref{eq:inifact}, it reads
\begin{equation}
d\mu( |\psi(0)\rangle)=\frac{1}{16\pi^2}\prod\limits_{k=1}^2 \sin\theta_k\text{d}\theta_k\text{d}\varphi_k.
\end{equation}
This measure is normalized to 1. It is trivial to see that in this case the entangling power $\mathfrak E$ lies within $[0,1]$.
Thus from \eqref{steadyconc} and \eqref{eq:enpower} we get 
\begin{equation}\label{epower}
\begin{aligned}
\mathfrak{E}=2\left|D\right|-2\sqrt{B_1 B_4}
+\delta_{\gamma,1}\delta_{\bar{n}_g,0}\dfrac{1}{4}.
\end{aligned}
\end{equation}
In Fig.\ref{fig:entpow3D0} it is shown the entangling power \eqref{epower} as a function of $\gamma$ and 
$\bar{n}_l$ for $\bar{n}_g=0$. There, we can see that for $\gamma<1$ and $\bar{n}_l=0$ it is $\mathfrak{E}=0$. In contrast, for any value of $\gamma>0.7$, there exists a nonzero optimal value of local thermal noise $\bar{n}_l$ maximizing the entangling power; a phenomenon reminiscent of \emph{stochastic resonance} effect \cite{Gammaitoni1998}. Such an optimal value of noise tends to increase as $\gamma$ approaches $1$ (as one can also argue from Fig.\ref{fig:entpowregions}).

\begin{figure}[ht!]
\centering
\includegraphics[width=0.5\textwidth]{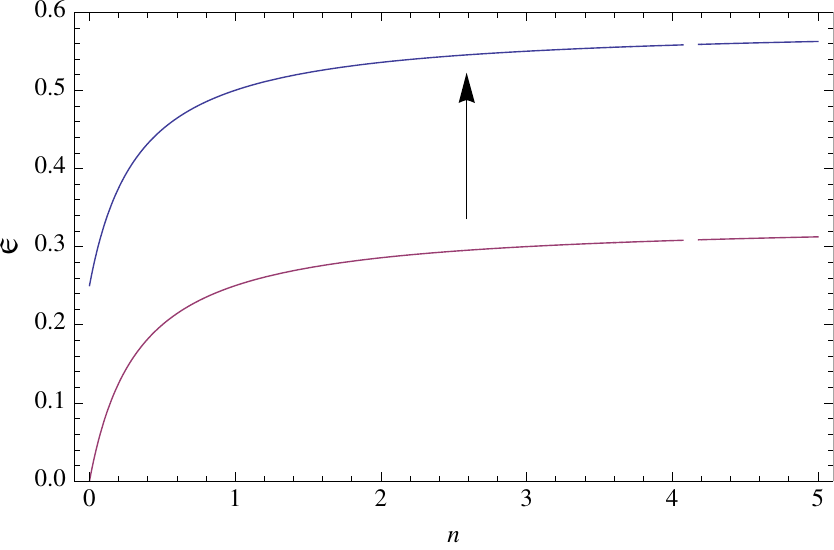}
    \fcaption{Entangling power $\mathfrak{E}$ as a function of  $\bar{n}_l$ 
    for $\bar{n}_g=0$ and $\gamma=1$. The bottom curve resulting from the contribution of the first two terms in  Eq.\eqref{epower} is shifted upward due to the contribution of the third term in Eq.\eqref{epower}.}
    \label{fig:entpowg1}
\end{figure}

When $\gamma$ attains the value 1, the curve of $\mathfrak{E}$ vs $\bar{n}_l$ is shifted upward by an amount $1/4$, as shown in Fig.\ref{fig:entpowg1}, and the maximum value of $\mathfrak{E}$, namely 7/12, is asymptotically achieved at $\bar{n}_l\to\infty$.

\begin{figure}[ht!]
\centering
\includegraphics[width=0.6\textwidth]{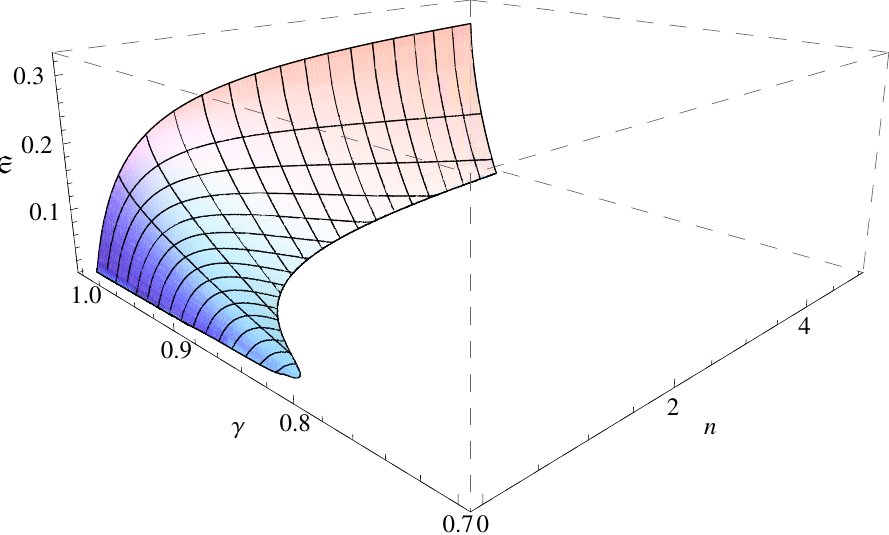}
    \fcaption{Entangling power $\mathfrak{E}$  as a function of $\gamma$ and $\bar{n}_l$ 
    for $\bar{n}_g=0.01$.}
    \label{fig:entpow3D001}
\end{figure}

\begin{figure}[ht!]
\centering
\includegraphics[width=0.6\textwidth]{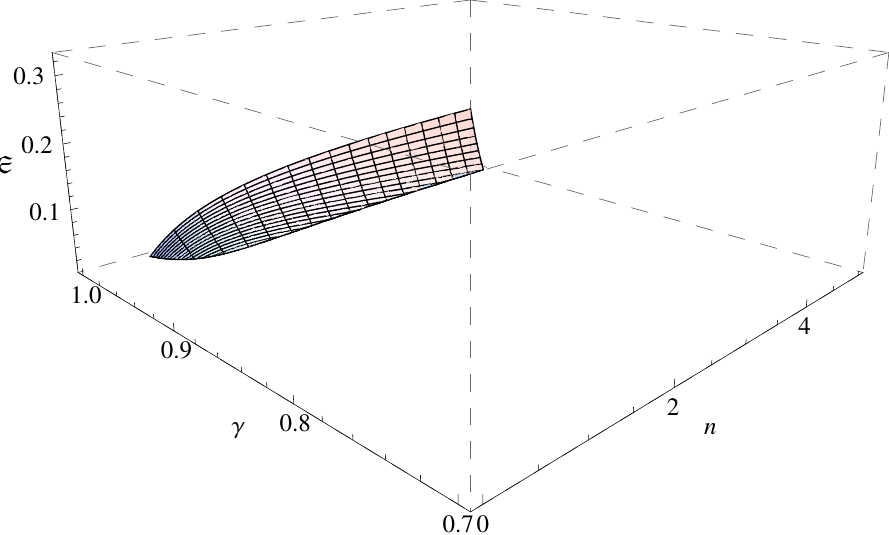}
    \fcaption{Entangling power $\mathfrak{E}$  as a function of $\gamma$ and $\bar{n}_l$ 
    for $\bar{n}_g=0.1$.}
    \label{fig:entpow3D01}
\end{figure}

By increasing the value of $\bar{n}_g$ from zero, the region $\{\gamma,\bar{n}_l\}$ of positive values of 
$\mathfrak{E}$ shrinks and also the maxima lower, as can be readily seen in Figs.\ref{fig:entpow3D001} and \ref{fig:entpow3D01}.
Hence we can conclude that global thermal noise is detrimental for stationary entanglement, while a suitable amount of local thermal noise is vital.

This can be explained by considering local thermal baths as injecting excitations onto the systems incoherently. Hence each excitation, thanks to the interaction mimicked by the global environment, is then shared by the two qubits ending up into an entangled state resembling $\frac{1}{\sqrt{2}}(|e_1g_2\rangle+|g_1e_2\rangle)$. If however the local noise is too much, it blurs such effect. In contrast, when $\bar{n}_g>0$ there is the tendency by the global bath to inject coherently two excitations onto the system which is then driven into a separable state resembling $|e_1 e_2\rangle$.

\section{Conclusions} \label{sec:conclusion}

In this paper, we have considered a model of two qubits dissipating by a ``glocal" map, i.e. into local and global environments (generally at finite temperatures). By the parameter $\gamma$, this can interpolate between perfect local and perfect global regimes. 

We have then determined conditions for the presence of long living entanglement.
This has been done by considering the entangling capabilities of the ``glocal" dissipative map ${\cal D}$ through the entangling power introduced in \eqref{eq:enpower}. 

It has been shown that the number of thermal excitations in the local environments has a crucial role on the stationary entanglement of the two qubits. 
 It results on the one hand (and rather counterintuitively) that entanglement in presence of local environments is achievable when these are at nonzero temperature.
This represents a remarkable extension of the \emph{stochastic resonance} effect 
arising in spin chains from the interplay of local dissipative and dephasing noise sources in the presence of Hamiltonian couplings \cite{Rivas2009}.
Here it appears in a context lacking of Hamiltonian couplings and driven by the interplay of the same kind (dissipative) of noise sources.
On the other hand, while global environment is vital for indirect qubits interaction, it should be ideally at zero temperature. In fact thermal noise from global environment spoils entanglement.
 
Concerning $\mathfrak{E}({\cal D})$ it is also worth noticing its sudden enhancement for zero temperature global dissipation (Fig. \ref{fig:entpowg1}). This can be regarded as a signature of a kind of phase transition occurring at $\gamma=1$.

The map ${\cal D}$, thanks to the properties discussed in Section \ref{sec:Kraus}, can also be considered
as a quantum channel and characterized in terms of its information transmission capabilities \cite{RMP14}. 
For instance, when $\gamma\neq 1$ the output space is of dimension 1, hence its capacity
vanishes. In contrast, when $\gamma = 1$ and $\bar{n}_g = 0$ the output space is of dimension 2 
(spanned by $\frac{1}{\sqrt{2}}\left(\ket{e_1}\ket{g_2}-\ket{g_1}\ket{e_2}\right)$ and $\ket{g_1}\ket{g_2}$) 
and the capacity could be up to 1 bit or 1 qubit
(depending on whether classical or quantum information is considered to be transmitted). 
For this rather different behavior in the parametric region, it could also be considered as a paradigmatic model for channel discrimination \cite{Pirs11}.
These investigations are left for future works.

The present study can be of interest for experimental situations where
the interplay of local and global environments is relevant.
As an example we may mention cavity QED experiments in which atomic qubits are confined inside a high finesse optical cavity  \cite{Guthahrlein2001} and experience local spontaneous emission as well as
a global effect of vacuum bath lying outside the cavity.
Another example is provided by charge qubits based on double quantum dots (or analogously Cooper pair boxes) \cite{CPA08} with local electron environments and global environment arising from voltage fluctuations.

\nonumsection{Acknowledgements}
\noindent
A.N. would like to thank the University of Camerino
for kind hospitality and the Ministry of Science, Research
and Technology of Iran for financial support.


\nonumsection{References}


\appendix{\;(Solution of Eq.\eqref{eq:differ} for $\gamma=1$ and $\bar{n}_g=0$)}\label{App:differ}

Taking $\gamma=1$ and $\bar{n}_g=0$ in Eq.(\ref{eq:differ}) and performing the inverse Laplace transform, one may find the following
matrix representing $\rho(t)$ with respect to the basis $\{\ket{1}, \ket{2}, \ket{3}, \ket{4}\}$:
\begin{equation}
\rho(t)= 
  \begin{pmatrix}
P_{11} & P_{12} \\
P_{21} & P_{22}
  \end{pmatrix},
\end{equation}
where 
\begin{equation}
P_{11}:= 
  \begin{pmatrix}
 A_1^2\rho_{11} & & A_7\rho_{12}+A_8\rho_{13} \\
 A_7\rho_{21}+A_8\rho_{31} & & A_2\rho_{11}+A_3\rho_{22}+A_4\rho_{33}+A_5(\rho_{23}+\rho_{32})
  \end{pmatrix},
\end{equation}

\begin{equation}
P_{12}:= 
  \begin{pmatrix}
A_8\rho_{12}+A_7\rho_{13} & & A_1\rho_{14} \\
A_2\rho_{11}+A_5(\rho_{22}+\rho_{33})+A_3\rho_{23}+A_4\rho_{32} & & -2A_8(\rho_{12}+\rho_{13})+A_9\rho_{24}+A_{10}\rho_{34} 
  \end{pmatrix},
\end{equation}

\begin{equation}
P_{21}:= 
  \begin{pmatrix}
A_8\rho_{21}+A_7\rho_{31} & & A_2\rho_{11}+A_5(\rho_{22}+\rho_{33})+A_4\rho_{23}+A_3\rho_{32}  \\
 A_1\rho_{41} & & -2A_8(\rho_{21}+\rho_{31})+A_{9}\rho_{42}+A_{10}\rho_{43}
  \end{pmatrix},
\end{equation}

\begin{equation}
P_{22}:= 
\begin{pmatrix}
A_2\rho_{11}+A_4\rho_{22}+A_3\rho_{33}+A_5(\rho_{23}+\rho_{32}) &  & -2A_8(\rho_{12}+\rho_{13})+A_{10}\rho_{24}+A_{9}\rho_{34} \\
-2A_8(\rho_{21}+\rho_{31})+A_{10}\rho_{42}+A_{9}\rho_{43} & & A_6\rho_{11}-2A_5(\rho_{22}+\rho_{23}+\rho_{32}+\rho_{33})+\rho_{44}
\end{pmatrix},
\end{equation}
and the time dependent coefficients $A_j$ are:
\begin{equation}
\begin{aligned}
A_1(t)&=e^{-2t},   \\
A_2(t)&=2 t e^{-4t},  \\
A_3(t)&=\frac{1}{4} e^{-4 t} \left(1+e^{2t}\right)^2,  \\
A_4(t)&=\frac{1}{4} e^{-4 t} \left(1-e^{2t}\right)^2, \\
A_5(t)&=-\frac{1}{4} \left(1-e^{-4t}\right),  \\
A_6(t)&=-e^{-4t}(1+4t-e^{4t}),  \\
A_7(t)&=\frac{1}{2} e^{-4 t} \left(1+e^{2t}\right),  \\
A_8(t)&=\frac{1}{4} e^{-4 t} \left(1-e^{2t}\right),  \\
A_9(t)&=\frac{1}{2} \left(1+e^{-2t}\right),  \\
A_{10}(t)&=-\frac{1}{2}\left(1-e^{-2t}\right).
\end{aligned}
\label{eq:coef}
\end{equation}
At the steady state (i.e., $t\to\infty$), we are left with only the following nonzero coefficients $A_3=A_4=-A_5=\frac{1}{4}$, $A_6=1$ and $A_9=-A_{10}=\frac{1}{2}$.  


\appendix{\;(Kraus operators of map \eqref{eq:dynmapt})} \label{App:Appendixcoeff}

The map \eqref{eq:dynmapt} can be represented as 
\begin{equation}
{\cal D}_t(\rho(0))=\sum_{j=1}^4K^{\prime\prime}_{j}(t)\rho(0) K^{\prime\prime\dagger}_{j}(t),
\end{equation}
where the (time dependent) Kraus operators can be derived by the spectral decomposition of the Choi matrix
associated to ${\cal D}_t$ \cite{Choi1972}, 
and given by $\left( {\cal I}\otimes{\cal D}_t\right) \left( \ket{\varPhi}\bra{\varPhi}\right) $, with $\ket{\varPhi}\equiv\ket{e_1}\ket{e_2}+\ket{e_1}\ket{g_2}+\ket{g_1}\ket{e_2}+\ket{g_1}\ket{g_2}$.
It explicitly results as
\setcounter{MaxMatrixCols}{20}
\begin{equation}
{\cal C}=\begin{pmatrix}
 A_1^2 & 0 & 0 & 0 & 0 & A_7 & A_8 & 0 & 0 & A_{8} & A_7 & 0 & 0 & 0 & 0 & A_1 \\
 0 & A_2 & A_2 & 0 & 0 & 0 & 0 & -2A_{8} & 0 & 0 & 0 & -2A_{8} & 0 & 0 & 0 & 0 \\
 0 & A_2 & A_2 & 0 & 0 & 0 & 0 & -2A_{8} & 0 & 0 & 0 & -2A_{8} & 0 & 0 & 0 & 0 \\
 0 & 0 & 0 & A_6 & 0 & 0 & 0 & 0 & 0 & 0 & 0 & 0 & 0 & 0 & 0 & 0 \\
 
 0 & 0 & 0 & 0 & 0 & 0 & 0 & 0 & 0 & 0 & 0 & 0 & 0 & 0 & 0 & 0 \\
A_7 & 0 & 0 & 0 & 0 & A_3 & A_5 & 0 & 0 & A_5 & A_3 & 0 & 0 & 0 & 0 & A_9\\
A_{8} & 0 & 0 & 0 & 0 & A_5 & A_4 & 0 & 0 & A_4 & A_5 & 0 & 0 & 0 & 0 & A_{10}\\
 0 & -2A_{8} & -2A_{8} & 0 & 0 & 0 & 0 & -2A_5 & 0 & 0 & 0 & -2A_5 & 0 & 0 & 0 & 0\\
 
 0 & 0 & 0 & 0 & 0 & 0 & 0 & 0 & 0 & 0 & 0 & 0 & 0 & 0 & 0 & 0 \\
 A_8 & 0 & 0 & 0 & 0 & A_5 & A_4 & 0 & 0 & A_4 & A_5 & 0 & 0 & 0 & 0 & A_{10} \\
 A_7 & 0 & 0 & 0 & 0 & A_3 & A_5 & 0 & 0 & A_5 & A_3 & 0 & 0 & 0 & 0 & A_9 \\
 0 & -2A_8 & -2A_8 & 0 & 0 & 0 & 0 & -2A_5 & 0 & 0 & 0 & -2A_5 & 0 & 0 & 0 & 0\\
 
0 & 0 & 0 & 0 & 0 & 0 & 0 & 0 & 0 & 0 & 0 & 0 & 0 & 0 & 0 & 0 \\
0 & 0 & 0 & 0 & 0 & 0 & 0 & 0 & 0 & 0 & 0 & 0 & 0 & 0 & 0 & 0 \\
0 & 0 & 0 & 0 & 0 & 0 & 0 & 0 & 0 & 0 & 0 & 0 & 0 & 0 & 0 & 0 \\
A_1 & 0 & 0 & 0 & 0 & A_9 & A_{10} & 0 & 0 & A_{10} & A_9 & 0 & 0 & 0 & 0 & 1 \\
 \end{pmatrix}.
 \label{eq:choimatrix}
\end{equation}
It is straightforward to show that this matrix is Hermitian and positive, therefore it can be expressed as the eigen-decomposition ${\cal C}=\sum_{j=1}^{4}\ket{c_j}\bra{c_j} $, with $\ket{c_j}$ its eigenvectors normalized to the respective eigenvalues. The $|c_j\rangle$ are then used to form the Kraus operators. Following \cite{Leung2003} this is done by dividing $\ket{c_j}$ (of length $4\times 4$) into 4 equal segments each of length 4. Then each Kraus operator $K^{\prime\prime}_j(t)$ is obtained by constructing a matrix having the $k$th segment as the its $k$th column.
They result as follow:
\begin{equation}
\begin{aligned}
K^{\prime\prime}_1(t)&=\begin{pmatrix}
 A_1(t) & 0 & 0 & 0 \\
  0 & A_9(t) & A_{10}(t) & 0 \\
 0 & A_{10}(t) & A_9(t) &  0 \\
 0 & 0 &  0 & 1
 \end{pmatrix}, \qquad\qquad
K^{\prime\prime}_2(t)=
\begin{pmatrix}
  0 & 0 & 0 & 0 \\
   0 & 0 & 0 & 0 \\
  0 & 0 & 0 &  0 \\
  \sqrt{A_6(t)} & 0 &  0 & 0
  \end{pmatrix}, \\
K^{\prime\prime}_3(t)&=\begin{pmatrix}
   0 & 0 & 0 & 0 \\
   \Lambda_+(t) & 0 & 0 & 0 \\
   \Lambda_+(t) & 0 & 0 &  0 \\
   0 & \Xi_+(t) & \Xi_+(t) & 0
   \end{pmatrix}, \qquad
K^{\prime\prime}_4(t)=\begin{pmatrix}
0 & 0 & 0 & 0 \\
   \Lambda_-(t) & 0 & 0 & 0 \\
   \Lambda_-(t) & 0 & 0 &  0 \\
   0 & \Xi_-(t) & \Xi_-(t) & 0
   \end{pmatrix},
\end{aligned}
\label{eq:kraus}
\end{equation}
where
 \begin{equation}
\begin{aligned}
\Lambda_{\pm}(t)&=\dfrac{\sqrt{2}(1-e^{-2t})}{\sqrt{\Upsilon(t)\left( \Upsilon(t)\pm e^{4t}A_6(t)\right) }}
\sqrt{ \Theta(t)\pm\Upsilon(t) } , \\
\Xi_{\pm}(t)&=\frac{e^{-2t}-1}{\sqrt{2}\Upsilon(t)\Lambda_{\pm}(t)}
\sqrt{\Theta(t)\pm\Upsilon(t)},
\end{aligned}
\end{equation}
and
 \begin{equation}
\begin{aligned}
\Theta(t)&:=-1+e^{4t}+4t, \\
\Upsilon(t)&:= 16 t^2-8 e^{4 t} t+8 t-32 e^{2 t}+14 e^{4 t}+e^{8 t}+17.
\end{aligned}
\end{equation}
Taking the limit $t\to\infty$ in Eq.\eqref{eq:kraus} we readily obtain Eq.\eqref{eq:krausinf}.
\end{document}